# Spatial correlation of two-dimensional Bosonic multi-mode condensates

Wolfgang H. Nitsche,[(a),[#1]] Na Young Kim,[(a)] Georgios Roumpos,[(a),[#2]] Christian Schneider,[(b)] Sven Höfling,[(b),(c),(d)] Alfred Forchel,[(b)] and Yoshihisa Yamamoto[(a),(d)]

[(a)] E. L. Ginzton Laboratory, Stanford University, Stanford, California 94305, USA
[(b)] Technische Physik, Universität Würzburg, Am Hubland, 97074 Würzburg, Germany
[(c)] School of Physics & Astronomy, University of St Andrews, St Andrews KY16 9SS, United Kingdom
[(d)] National Institute of Informatics, Hitosubashi, Chiyoda-ku, Tokyo 101-8430, Japan

We studied the spatial coherence of a Bosonic two-dimensional multi-mode condensate both through measurements and simulations. It is shown that condensates with a constant spatial density must be described as the superposition of several quantized modes which reduces the overall coherence. In this case, the spatial coherence can appear to decay faster than allowed by the Berezinskii-Kosterlitz-Thouless (BKT) theory. However, we find through spectroscopic measurements that the individual modes show a slower decay of the spatial coherence than the overall system.

***Introduction***. - Two-dimensional Bosonic systems at low temperature have been observed to exhibit spatial coherence over long distances [Deng2007] [Hadzibabic2006] [Kasprzak2006] [Nitsche2014] [Roumpos2012]. Although Bose-Einstein condensation with true long-range order is not possible in such systems [Hohenberg1967] [Mermin1966], they can still exhibit quasi-long-range order by condensing into a state which is described by the Berezinskii-Kosterlitz-Thouless (BKT) theory [Berezinskiĭ1972] [Kosterlitz1973]. In this case, the first-order spatial correlation function is predicted to decay with a power-law (algebraically), whose exponent must always be $\leq 1/4$ [Hadzibabic2011]. Here, we show that the spatial quasi-long-range-order of a condensate can decay faster than allowed under the BKT theory, if the system shows simultaneous condensation into several spatio-energetic modes. The Nozieres principle [Nozières1995] in general predicts that condensation will only occur into a single mode since condensate fragmentation is energetically costly due to the Fock exchange term; however multi-mode condensation is possible in a system where the energy uncertainty caused by the short Boson lifetime is greater than the Fock exchange energy. Here we confirm the fast decay of the spatial coherence in a multi-mode condensate both by numerical simulations and through experimental measurements using an exciton-polariton system.

***Bosonic condensate with constant density***. - We study a finite sized two-dimensional Bosonic condensate with a nearly constant particle density $n_{\text{total}}(x,y)$ throughout the condensate. In this paper, we focus on the case

$$n_{\text{total}}(x,y) \approx n_{\text{const}} H\left(R_{\text{spot}} - \sqrt{x^2+y^2}\right), \tag{0}$$

where $H$ is the Heaviside step function. This corresponds to the most relevant case of a condensate occupying a circular area with radius $R_{\text{spot}}$; but condensates with rectangular or other forms could be studied in an analog way. Such a condensate with a constant density cannot be in a single quantum mechanical mode, since this would require a wave-function whose amplitude is constant within the circular area and zero everywhere else, but such a wave-function is not an eigenfunction to the Schrödinger equation. However a nearly constant total density is possible if condensation occurs simultaneously into several spatio-energetic modes $\Psi_s$ which gives [#3] a total wave-function of the form

$$\Psi_{\text{total}}(x,y,t) = \sum_s \Psi_s(x,y,t), \tag{1}$$

where a mode number $s$ is described by

$$\Psi_s(x,y,t) = A_s \psi_s(x,y) e^{itE_s/\hbar} e^{i\eta_s(x,y,t)}. \tag{2}$$

Due to the circular symmetry of the condensate, each base-function $\psi_s$ can be expressed [Roumpos2010] (in polar coordinates) as

$$\psi_s(r,\phi) = J_{m_s}(k_s r)\exp(im_s\phi) H(R_{\text{spot}} - r), \tag{3}$$

where the orbital quantum number $m_s$ is an integer, $J$ is the Bessel function of the first kind, the positive $k_s$ is chosen so that the Bessel function reaches one of its zeros at $r = R_{\text{spot}}$ and $E_s$ is the energy corresponding to base-function $\psi_s$. The amplitudes $A_s$ are determined by the condition

$$\left\langle \Psi_{\text{total}}(x,y,t) \Psi^{\dagger}_{\text{total}}(x,y,t) \right\rangle = n_{\text{total}}(x,y). \tag{4}$$

The real $\eta_s(x,y,t)$ represents phase noise and can for example be explained by the thermal excitation [Hadzibabic2011] of phononic long-wavelength phase fluctuations. Without this phase-noise, coherence could exist over arbitrary large distances which is not possible [Hohenberg1967] [Mermin1996] in our two-dimensional system. The $\langle \ldots \rangle$ brackets indicate time-averaging.

***Correlation function of our system***. - The first order spatial correlation function can be defined as

$$g^{(1)}(x_1,y_1,x_2,y_2) = \frac{\left|\left\langle \Psi(x_1,y_1,t)\Psi^{\dagger}(x_2,y_2,t)\right\rangle\right|}{\sqrt{\left\langle \left|\Psi(x_1,y_1,t)\right|^2\right\rangle \left\langle \left|\Psi(x_2,y_2,t)\right|^2\right\rangle}}, \tag{5}$$

and it reaches a value of 1 if there is perfect coherence between points $(x_1,y_1)$ and $(x_2,y_2)$, but its value is 0 if there is no coherence between these points. The coherence for each individual mode $\Psi_s$ can therefore be written as

$$\begin{aligned} g_s^{(1)}(x_1,y_1,x_2,y_2) &= \frac{\left|\left\langle \Psi_s(x_1,y_1,t)\Psi_s^{\dagger}(x_2,y_2,t)\right\rangle\right|}{\sqrt{\left\langle \left|\Psi_s(x_1,y_1,t)\right|^2\right\rangle \left\langle \left|\Psi_s(x_2,y_2,t)\right|^2\right\rangle}} = \\ &= \left|\left\langle \exp\left\{i\left[\eta(x_1,y_1,t) - \eta(x_2,y_2,t)\right]\right\}\right\rangle\right|. \end{aligned} \tag{6}$$

It can be shown [Hadzibabic2011] that for condensation into a single mode $\Psi_s$, this coherence function over large distances behaves like

$$g_s^{(1)}(x_1,y_1,x_2,y_2) \propto \sqrt{(x_1-x_2)^2+(y_1-y_2)^2}^{\,-a_\mathrm{p}}, \tag{7}$$

with $a_\mathrm{p} \le 1/4$. For short distances, it is expected to converge towards 1. The coherence of the total system $\Psi_\mathrm{total}$ can be calculated (using equations (1), (2), and (6)) as

$$\begin{aligned} g_\mathrm{total}^{(1)}(x_1,\mathrm{y}_1,x_2,y_2) &= \frac{\left|\left\langle \Psi_\mathrm{total}(x_1,y_1,t)\Psi_\mathrm{total}^\dagger(x_2,y_2,t)\right\rangle\right|}{\sqrt{\left\langle\left|\Psi_\mathrm{total}(x_1,y_1,t)\right|^2\right\rangle\left\langle\left|\Psi_\mathrm{total}(x_2,y_2,t)\right|^2\right\rangle}} = \\ &= \frac{\left|\sum_s |A_s|^2 \psi_s(x_1,y_1)\psi_s^\dagger(x_2,y_2)\,g_s^{(1)}(x_1,\mathrm{y}_1,x_2,y_2)\right|}{\sqrt{\left(\sum_s |A_s\psi_s(x_1,y_1)|^2\right)\left(\sum_s |A_s\psi_s(x_2,y_2)|^2\right)}}. \end{aligned} \tag{8}$$

Here we used

$$\begin{aligned} &\left\langle \exp\left\{i\left[\eta_{s_1}(x_1,y_1,t)-\eta_{s_2}(x_2,y_2,t)\right]\right\}\exp\left(i\frac{E_{s_1}-E_{s_2}}{\hbar}t\right)\right\rangle = \\ &= \left\langle \exp\left\{i\left[\eta_{s_1}(x_1,y_1,t)-\eta_{s_2}(x_2,y_2,t)\right]\right\}\right\rangle \delta(s_1,s_2), \end{aligned} \tag{9}$$

which assumes that the energies are non-degenerate, so that for $s_1 \ne s_2$ all interference terms with $\exp\left[i\left(E_{s_1}-E_{s_2}\right)t/\hbar\right]$ average out to zero [#4]. Equation (8) shows that even if each individual mode is perfectly coherent ($g_s^{(1)} \approx 1$), the overall coherence $g_\mathrm{total}^{(1)}$ can nevertheless exhibit a fast decay, as a direct result of the simultaneous excitation of several modes $\Psi_s$. The details of which modes $\Psi_s$ are excited are determined by the spatial particle density $n_\mathrm{total}(x,y)$ with which the condensate is created. This also implies that if we create a condensate having a (non-constant) particle density $n_\mathrm{total}(x,y)=\left|\Psi_{\tilde{s}}(x,y)\right|^2$ which comprises only one chosen mode $\Psi_{\tilde{s}}$, we get $A_s = 0$ for $s \ne \tilde{s}$, but $A_{\tilde{s}} \ne 0$, and the coherence $g_\mathrm{total}^{(1)}$ becomes identical to the coherence $g_{\tilde{s}}^{(1)}$ of the chosen mode.

**Michelson interferometer.** - The correlation function and phase of a condensate could for example be determined with a Michelson interferometer which flips the signal in one arm around the $y$-axis, so that effectively $\Psi_\mathrm{total}(x,y)$ interferes with $\Psi_\mathrm{total}(-x,y)$ as explained in previous publications [Nitsche2014] [Roumpos2012]. The measured intensity (as a function of the path length difference $\Delta L$ between both arms of the Michelson interferometer)

$$I_\mathrm{measured} = I_B + I_A \sin\left(w\Delta L - \varphi_0\right), \tag{10}$$

tells us the phase $\varphi_0$ and the visibility $g^{(1)} = I_A/I_B$ for each point. The average intensity $I_B$ is approximately proportional to the particle density. We assume that the path length difference is always sufficiently small so that issues related to temporal coherence and similar aspects do not affect the measurement.

***Simulated data***. - For the simulation, we choose a few possible modes $\Psi_s$ and fit the amplitudes $A_s$ so that $n_{\text{total}}(x,y)$ is approximately proportional to $H\left(10\,\mu\text{m}-\sqrt{x^2+y^2}\right)$, as shown in Fig. 1(a,b). The intensity (Fig. 1(c)) of the Michelson interference image as it would be measured is

$$
\begin{aligned}
I_{\text{interference}}(x,y) &\propto \left\langle \Psi_{\text{total}}(x,y,t)\Psi^{\dagger}_{\text{total}}(-x,y,t)e^{-i\alpha(x)}\right\rangle = \\
&= \sum_s |A_s|^2 \left\{ |\psi_s(x,y)|^2 + |\psi_s(-x,y)|^2 + 2g_s^{(1)}(x,y,-x,y)\Re\left[\psi_s(x,y)\psi_s^{\dagger}(-x,y)e^{-i\alpha(x)}\right]\right\}
\end{aligned}
\tag{11}
$$

where the phase difference $\alpha(x)\approx w\Delta L + vx$ is determined by the path length difference $\Delta L$ and the different angle with which the signal from both arms reach the plane where the intensity is measured [#5]. Even with the assumption of $g_s^{(1)}\approx 1$ (which we used for the simulations since the predicted slow algebraic decay is negligible over the small condensate size), the $g^{(1)}_{\text{total}}(x,y,-x,y)$ which we simulated using equation (8) shows a clear decay (Fig. 1(f)). The simulated interference signal (Fig. 1(c)) and the simulated interference phase (Fig. 1(d)) show forks in the fringes which look similar to trapped vortices [Manni2013], but in our case, where we did not assume any trapped vortices, they are an artifact caused by the simultaneous presence of multiple modes $\Psi_s$.

***Experiments with exciton-polaritons***. - To confirm our theoretically derived predictions about the spatial coherence of a two-dimensional Bosonic condensate, we performed measurements with exciton-polaritons. Exciton-polaritons are a quantum mechanical superposition of quantum-well (QW) excitons with cavity photons [Weisbuch1992], and they behave as Bosonic quasi-particles. At low temperatures, they show condensation [Deng2002] [Kasprzak2006] accompanied by spatial coherence [Deng2007] [Kasprzak2006] [Nitsche2014] [Roumpos2012]. The photons which leak out of the sample during the decay of the exciton-polaritons can easily be measured and they preserve the energy, in-plane-momentum and coherence properties of the decaying exciton-polaritons. Therefore, we perform Michelson interferometry of the leaked photons; this has been done with the setup described in previous publications [Nitsche2014] [Roumpos2012], and is much easier than performing direct of interference of an actual Bosonic condensate with itself. This measurement allows us to determine the first-order spatial coherence function as well as the interference phase. The sample used for these measurements (same as in previous experiments [Nitsche2014]) consists of 4 GaAs QWs in an AlAs $\lambda/2$ cavity which is confined on both sides by AlAs/AlGaAs Bragg reflectors. A helium flow cryostat is used to keep it at low temperature. The sample is non-resonantly excited by perpendicular incident continuous-wave laser-light at a wavelength which coincides with a reflection minimum of the Bragg reflector. The laser is chopped at 100 Hz with a duty-cycle of 5% to reduce thermal heating. The incident laser beam initially has a Gaussian spatial profile, but for most measurements an additional beam-shaper is used to change it to a top-hat form. The top-hat form excitation is intended to create an exciton-polariton condensate with a constant density throughout the area where the laser spot excites the sample [Roumpos2010], and no exciton-polaritons outside of this area.

***Spatially resolved measurement***. - Measured data (Fig. 2(a-f)) with a top-hat excitation of the sample behaves as the simulation (Fig. 1(d-f)) predicts: We see a fast decay of the spatial coherence (Fig. 2(c,f)), and "pseudo-vortices" are visible as forks in the phase (Fig. 2(a,d)). The measurement shown in Fig. 2(a-c) has been performed at low temperature where coherence is the result of exciton-polarton condensation. However the data for Fig. 2(d-f) has been measured using the same sample at a higher temperature of around 200 K where strong coupling is lost; thus, we attribute the evolution of spatial coherence to conventional (VCSEL) lasing in the weak coupling regime. The "pseudo-vortices" appear in both cases (Fig. 2(a,d)). This indicates that (as also the simulation in Fig. 1(d) showed) they are not related to real vortices in a BKT condensate, but rather they are artifacts from simultaneous emission at different energies $E_s$, so that it does not even matter if $\Psi_s$ represents a BKT condensate or a VCSEL mode. If we slightly move the sample, the "pseudo-vortices" do not move with the sample, but stay at a fixed position relative to the pump-beam. This also shows that they cannot be real vortices trapped at defects of the sample. For comparison, we also performed similar measurements [Nitsche2014] with a Gaussian spatial pump profile, which has only a non-negligible overlap with the lowest mode $\Psi_1$. In this case, condensation occurs only into this mode (Fig. 3(c)), and as expected, quasi-long-range order which only decays very slowly has been observed, and no "pseudo-vortices" appeared (Fig. 2(g-i)).

**Spectroscopic energy resolved measurement**. - Finally, to confirm that the simultaneous population of multiple modes reduces the spatial correlation, we performed an energy resolved measurement which allows us to measure the correlation function of the individual modes (Fig. 4). The setup is again similar to the previously used one [Nitsche2014] [Roumpos2012], but this time the interference image is projected onto the entrance slit of a spectrometer (Fig. 4(a)). If this slit is widely opened and the grating aligned to act as a mirror (0th order), the spectrometer CCD camera records position $x$ and $y$ resolved interference images (Fig. 4(b)) in which we see interference fringes. However the spectrometer can also be used to perform position $x$ and energy $E$ resolved measurements by closing the entrance slit so that only the signal with $y \approx 0$ enters the spectrometer, and aligning the grating in such a way (1st order) that the one-dimensional interference signal which passes through the slit becomes spectrally resolved and is recorded as an $x$ and $E$ resolved image (Fig. 4(c)). The data in this image has been recorded with one specific path-length difference $\Delta L$, but the same kind of measurement has also been performed for many different path-length differences. In this spectrographic data, we can identify at least 10 individual modes at different energy levels. The 11 horizontal red lines show the (manually determined) borders between individual modes; this means for example we assume that any signal in the energy interval labeled 6 (which is between the 6th and 7th red lines) corresponds to the sixth quantized mode. Within each individual mode, interference fringes are again obvious, and if the measurement is repeated with changing path-length differences, these fringes move in the horizontal $x$ direction. Fig. 4(d) depict the evaluation for the 6th mode (but exactly the same evaluation has also been performed for all the other modes). The black line is determined as $I_{\text{measured}}^{(\text{mode } 6)}(x) = \int_{E_6}^{E_7} I_{\text{measured}}(x, E)\,\mathrm{d}E$ where $E_6$ and $E_7$ are the energy levels at which we drew the 6th and 7th lines in Fig. 4(c) and $I_{\text{measured}}(x,E)$ is the measured spectrally resolved data in Fig. 4(c). This means the black line shown in Fig. 4(d) corresponds to one specific path-length difference in the same way as the data shown in Fig. 4(c). The clear oscillations in $I_{\text{measured}}^{(\text{mode }6)}(x)$ correspond again to the interference fringes, and they appear to move in the horizontal $x$ direction if one changes the path-length difference $\Delta L$. By performing the same integration for multiple path-length differences we get $I_{\text{measured}}^{(\text{mode }6)}(x,\Delta L)$, which is not shown in the figure, and by performing the sine fit explained in equation (10), we determine $I_A^{(\text{mode }6)}(x)$ and $I_B^{(\text{mode }6)}(x)$ as well as the (not shown) phase $\varphi_0^{(\text{mode }6)}(x)$. In Fig. 4(e), we show the visibilities which we calculate for each of the 10 individual modes as $g_{(\text{mode }s)}^{(1)}(x,-x)=I_A^{(\text{mode }s)}(x)/I_B^{(\text{mode }s)}(x)$. The continuous black line in Figure 6(f) displays the overall visibility which has been determined from the position resolved measurement where only data within the area between the black lines in Fig. 6(b) has been considered. Since this data has been measured without spectral energy resolving, it is equivalent to the data shown in Fig. 2(c). The dotted blue line in Fig. 4(f) has been calculated by undoing the energy resolution. For this, we did the same evaluation as in Fig. 4(d), but we started by integrating over all energies from $E_1$ to $E_{11}$. We notice that the dotted blue line which we get this way is approximately the same as the continuous black line which we got without using any spectroscopy. This confirms that we can undo the spectroscopy by integrating over the energy of the spectroscopic data. For comparison, the dashed green line shows the average of the correlation of the individual modes, weighted by their intensities, this means

$$g_{\text{weighted average}}^{(1)}(x,-x)=\frac{\sum_{s=1}^{10} I_A^{(\text{mode }s)}(x)\, g_{(\text{mode }s)}^{(1)}(x,-x)}{\sum_{s=1}^{10} I_A^{(\text{mode }s)}(x)}. \tag{12}$$

As expected, these weighted average which we get from evaluating individual modes is more coherent than the correlation which has been determined without energy resolution. Finally, we use the correlations, intensities and phases of the individual modes to calculate a phase sensitive average (continuous red line) using

$$g_{\text{phase sensitive average}}^{(1)}(x,-x)=\frac{\left|\sum_{s=1}^{10} I_A^{(\text{mode }s)}(x)\, g_{(\text{mode }s)}^{(1)}(x,-x)\exp\left(i\varphi_0^{(\text{mode }s)}\right)\right|}{\sum_{s=1}^{10} I_A^{(\text{mode }s)}(x)} \tag{13}$$

which follows from equation (8); as expected it fits the non-energy-resolved data which confirms our theory.

***Conclusion***. - In conclusion, a BKT condensate with constant spatial density must consist of several modes, the simultaneous presence of which causes a faster than expected decay of the spatial coherence and the appearance of "pseudo-vortices" which look similar to pinned vortices [Lagoudakis2008] [Lagoudakis2009] [Lagoudakis2011] [Marchetti2010] or a vortex lattice [Manni2013] in a BKT phase. The higher spatial coherence of individual modes has been recovered through energy resolved measurements. The simultaneous condensation into multiple modes might also explain the previously reported [Roumpos2012] decay of the coherence of an exciton-polariton condensate with a power-law whose exponent is larger than allowed under the BKT theory.

***Acknowledgement***. - This research has been supported by the Japan Society for the Promotion of Science (JSPS) through its "Funding Program for World-Leading Innovative R&D on Science and Technology (FIRST Program)", by Navy/SPAWAR Grant N66001-09-1-2024, and by National Science Foundation ECCS-09 25549. W. H. N. acknowledges the Gerhard Casper Stanford Graduate Fellowship.

***Footnotes and References***.

[#1] Corresponding author: http://www.nitsche.mobi nitsche@stanford.edu present address: Halliburton, Houston, Texas 77032, USA

[#2] present address: Google, Mountain View, California 94043, USA

[#3] Here, we assume a condensation fraction of nearly 100%; if we consider the presence of non-condensed Bosons we can perform similar calculations but receive a slightly lower spatial coherence.

[#4] If the camera which records the interference image has an integration time of $\Delta t = 1\,\mathrm{s}$, this assumption will be true as long as the energy difference between different states is larger than $\Delta E = h/\Delta t \approx 6.63\times10^{-34}\,\mathrm{J} = 4.14\times10^{-15}\,\mathrm{eV}$. Even for two modes as defined in equation (3) which differ only by the sign of $m_s$, under real experimental conditions there will always be a sufficient symmetry breaking which causes an energy difference of more than this $\Delta E$.

[#5] The constant $w \approx E/(\hbar c)$ depends on the energy $E$ of the photons emitted by the condensate. Technically, each mode $\Psi_s$ has a slightly different energy $E_s$ of the photons emitted by the condensate, but we can neglect this (especially for $\left|E_{s_{\min}} - E_{s_{\max}}\right| \Delta l << \hbar c$). In the same way, we may also assume that $v$ is constant.

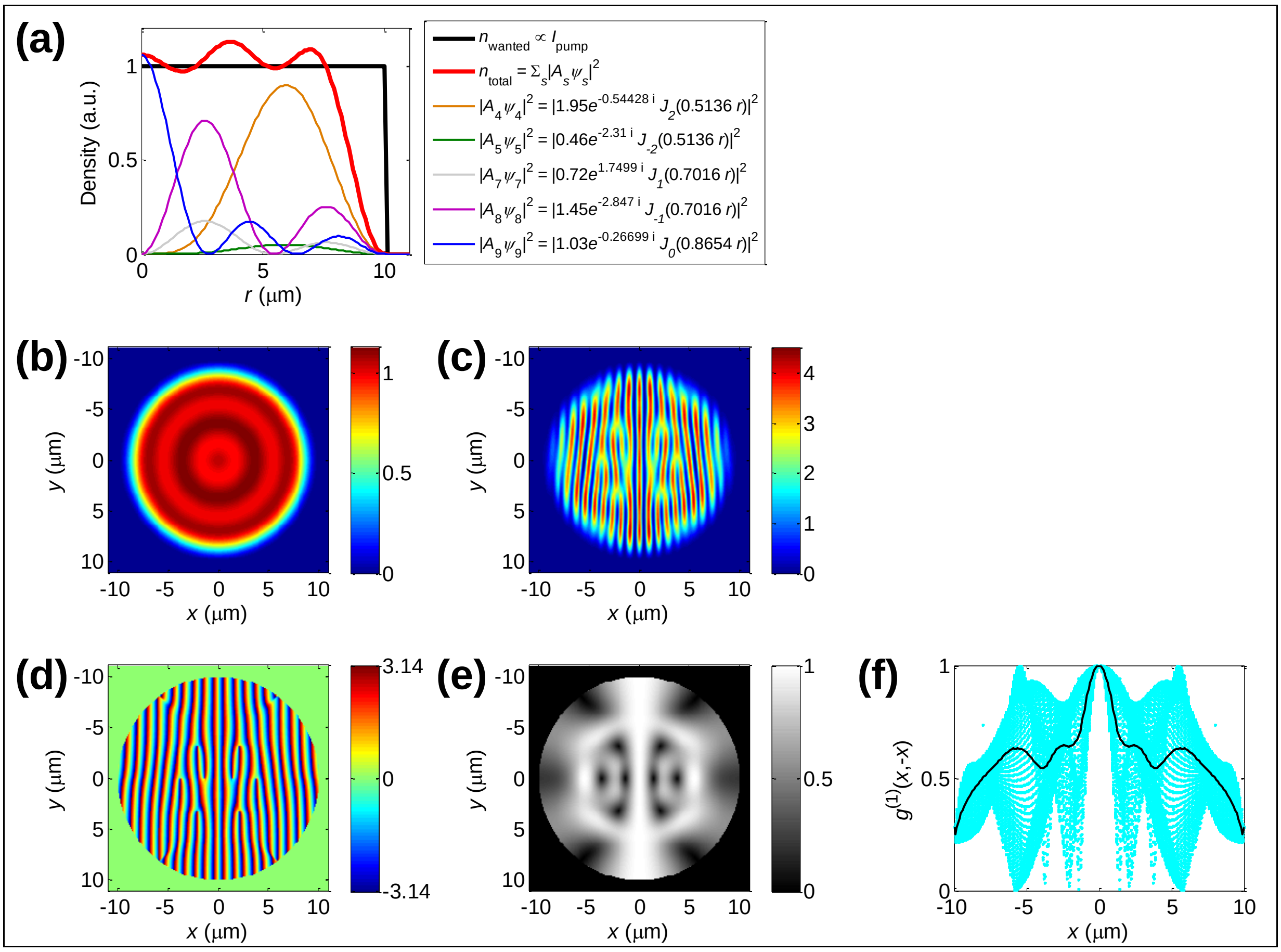

**FIG. 1: Simulated data.**

(a) A total density $n_{\text{total}}(r)$ compared to the $n_{\text{wanted}}(r) = H(10\,\mu\text{m} - r) \times 1\,\text{a.u.}$ . The contributions $\left|A_s \Psi_s(r,\phi)\right|^2$ of the individual modes are also shown.

(b) A total density $n_{\text{total}}(x, y)$.

(c) Interference signal as it would be measured by the camera of a Michelson interference setup.

(d) Phase as it would be determined through evaluating interference data.

(e) A correlation function $g^{(1)}_{\text{total}}(x, y, -x, y)$ as it would be calculated from the interference data.

(f) All the values of (e), but plotted as a function of $x$. Each cyan point corresponds to one pixel of (e). The continuous black line shows the averages of all the cyan data points with the same $x$-value. The oscillatory behavior can be understood if we assume that one of the higher modes dominates at larger distances, since the lowest mode is only significant in the region close to the center.

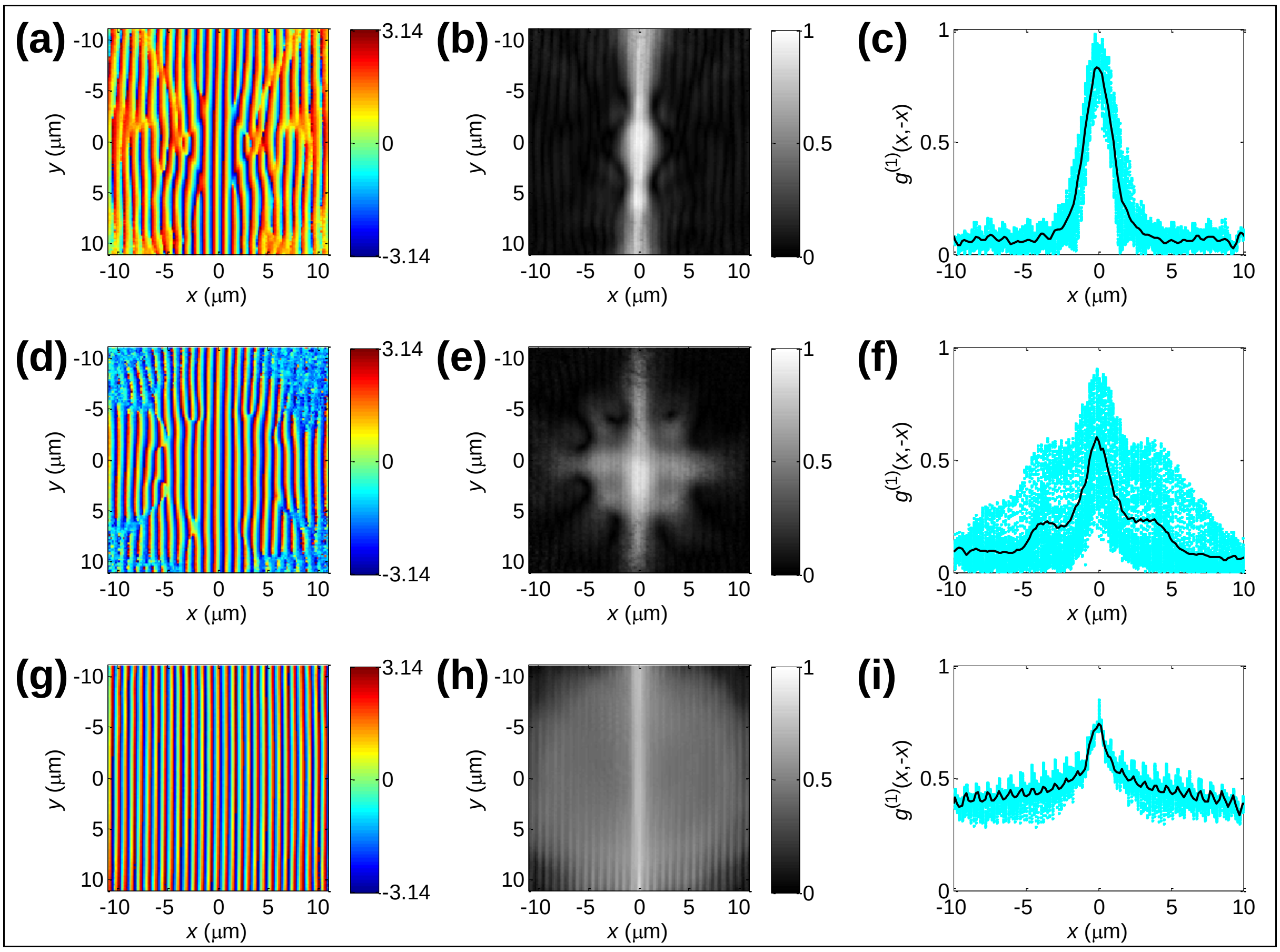

**FIG. 2: Measured data.**

(a-c) Measured phase (a) and visibility (b,c) of an exciton-polariton condensate with nearly constant density which has been created by pumping the sample with a top-hat profile at low temperature ($\approx 7$ K). The decay of the visibility is significantly faster than measured (i) for a single-mode BKT condensate [Nitsche2014] with quasi-long-range order. Pseudo-vortices which appear as forks in (a) are clearly visible. Each cyan point in (c) corresponds to one pixel of (b), and the continuous black line in (c) has been calculated by averaging all the cyan data points with the same $x$-value.

(d-e) The same kind of measurement (still with a top-hat pump profile) at $\approx 200$ K where the coherence can only be explained by VCSEL lasing. Nevertheless, the behavior is the same as for (a-c), which confirms that the visibility is determined by the presence of several quantized modes, rather than by the intrinsic coherence properties of a condensate.

(g-h) The same kind of measurement at low temperature ($\approx 5$ K) with a Gaussian pump profile which excites a single-mode exciton-polariton condensate. No pseudo-vortices (forks) appear in the phase map (g).

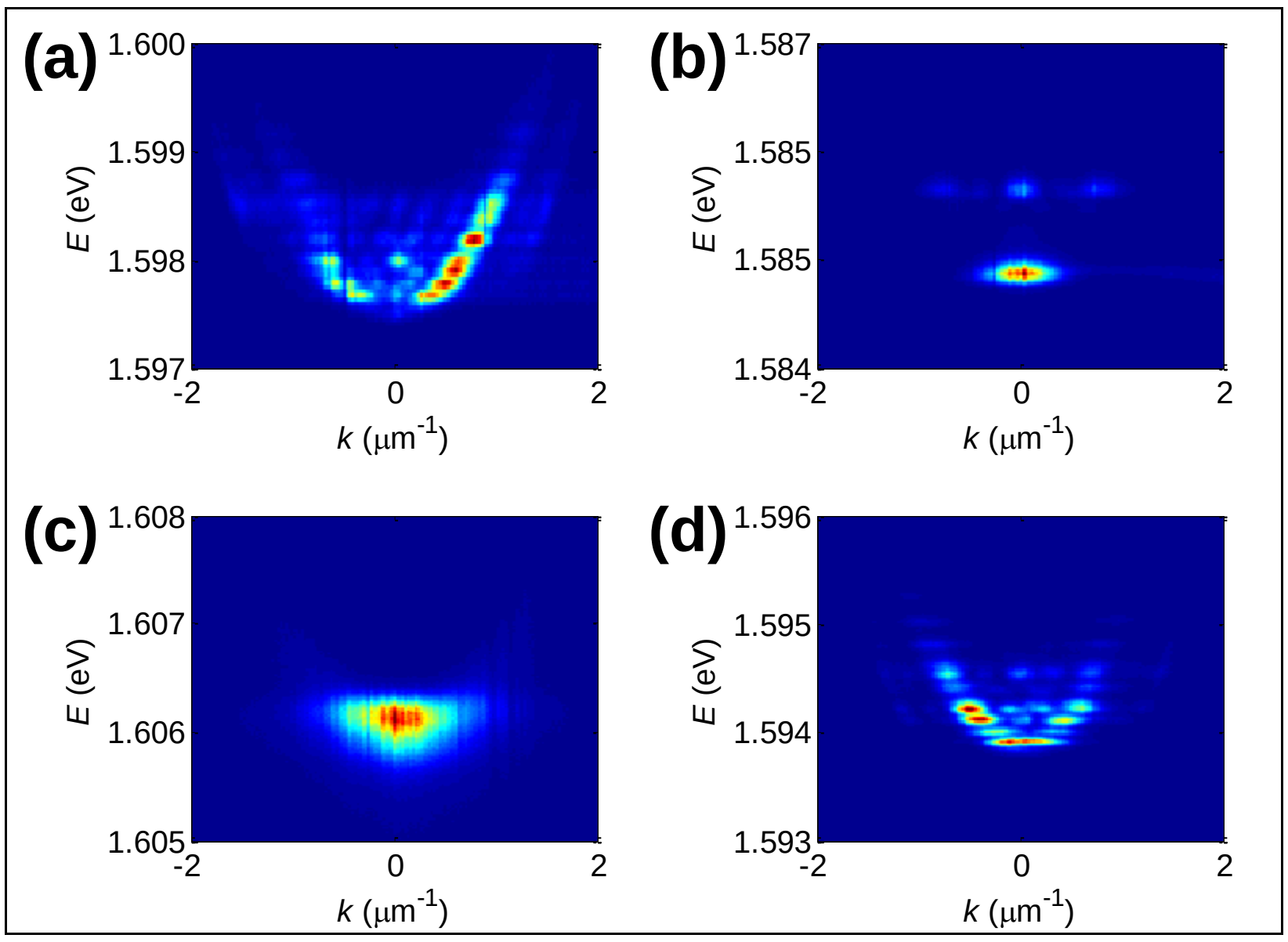

**FIG. 3: Measured dispersion.**

(a) Same experimental conditions as Fig. 2(a-c) (exciton-polariton condensation with top-hat pump profile at $\approx 7$ K ). Many quantized modes at different energy levels are clearly visible.

(b) Same experimental conditions as Fig. 2(d-f) (VCSEL lasing with top-hat pump profile at $\approx 200$ K ). At least two quantized modes at different energies can be seen.

(c) Same experimental conditions as Fig. 2(g-i) (exciton-polariton condensate with Gaussian pump profile at $\approx 5$ K ). Condensation occurs in a single mode at one energy level.

(d) Same experimental conditions as Fig. 4 (exciton-polariton condensation with top-hat pump profile at $\approx 8$ K ). Many quantized modes at different energy levels are clearly visible, and the modes are sufficiently energetically separated to allow the selection of individual modes through spectroscopy as shown in Fig. 4.

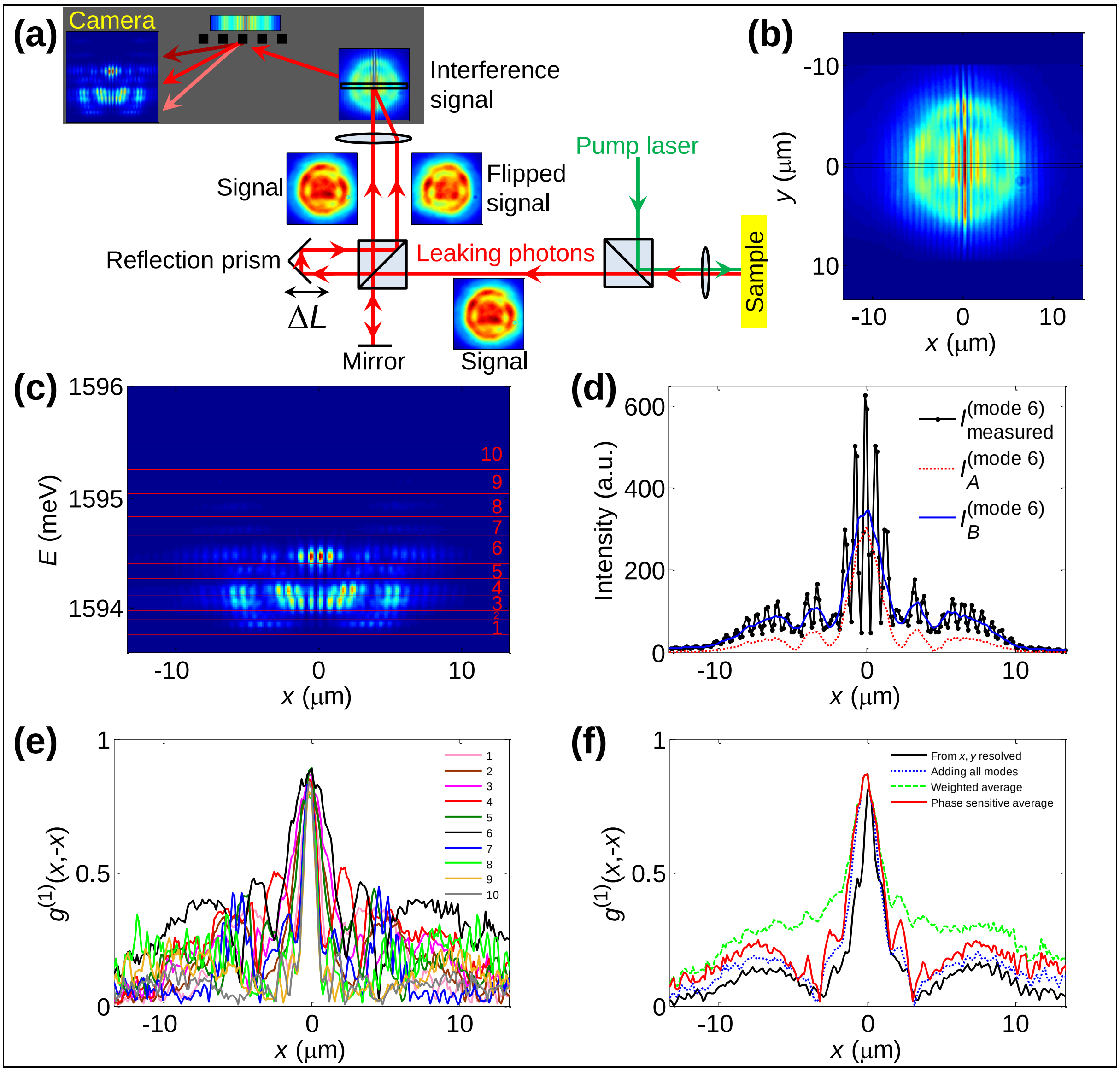

**FIG. 4: Spectrometric resolved interference measurement (exciton polariton condensate with nearly constant density, created at $\approx 8\,\mathrm{K}$ with top-hat pumping). The estimated detuning was -8 meV which is far red detuned.**

(a) A spectroscopic interference setup.

(b) A position $x$ and $y$ resolved interference image (for one specific path length difference $\Delta L$).

(c) A position $x$ and energy $E$ resolved interference image, again for one specific path length difference. Here, we can identify at least 10 individual modes at different energy levels.

(d) Evaluation for the 6th mode. $I_{\mathrm{measured}}^{(\mathrm{mode}\ 6)}$ has been measured at one specific path length difference, whereas $I_A^{(\mathrm{mode}\ 6)}$ and $I_B^{(\mathrm{mode}\ 6)}$ have been calculated by using the measured $I_{\mathrm{measured}}^{(\mathrm{mode}\ 6)}$ from many different path length differences.

(e) Visibilities $g_{(\mathrm{mode}\ s)}^{(1)} = I_A^{(\mathrm{mode}\ s)} / I_B^{(\mathrm{mode}\ s)}$ for the 10 individual modes.

(f) The overall correlation (black line) which has been calculated from the position $x$ and $y$ resolved measurement where only data within the area between the black lines in (b) has been considered. For comparison we also show different approaches to calculate the overall correlation from the position $x$ and energy $E$ resolved interference image (c).